\documentclass[journal, 10pt]{IEEEtran}
\usepackage{}

\usepackage{bm}
\usepackage{amsmath,amssymb}
\usepackage{graphicx,subfigure}
\usepackage{float}

\usepackage{color}

\definecolor{blue}{rgb}{0, 0, 1}

\newcommand{\Proj}{\mathbb{P}}
\newcommand{\EE}{\mathcal{E}}
\newcommand{\VV}{\mathcal{V}}
\newcommand{\RR}{\mathbb{R}}
\newcommand{\Sm}{\mathbb{S}}

\newtheorem{Assumption}{Assumption}
\newtheorem{Remark}{Remark}
\newtheorem{Lemma}{Lemma}

\newtheorem{Theorem}{Theorem}

\newtheorem{Problem}{Problem}
\newtheorem{Definition}{Definition}
\newenvironment{Proof}{\noindent{\em Proof:\/}}{\hfill $\Box$\par}

\begin{document}

\title{Bearing-based  Formation with Disturbance Rejection}

\author{Haoshu~Cheng and~Jie~Huang, \IEEEmembership{Life Fellow, IEEE}
\thanks{
\emph{(Corresponding author: Jie Huang.)}}
\thanks{This work was supported in part by the Research Grants Council of the Hong Kong Special Administration Region under grant No. 14201420, and in part by National Natural Science Foundation of China under Project 61973260.}
\thanks{H. Cheng and J. Huang  are with the Department of Mechanical and Automation Engineering,
The Chinese University of Hong Kong, Shatin, N.T., Hong Kong.
E-mail: hscheng@link.cuhk.edu.hk, jhuang@mae.cuhk.edu.hk}}

\maketitle

\begin{abstract}
This paper considers the problem  of the bearing-based formation control with disturbance rejection for a group of agents under the leader-follower structure.
The disturbances are in the form of a trigonometric polynomial with arbitrary unknown amplitudes, unknown initial phases, and known or unknown frequencies. For the case of
the known frequencies, we employ the canonical internal model to solve the problem, and,  for the case of
the unknown frequencies,  we combine the canonical internal model and {some} distributed adaptive control technique to deal with the problem.
It is noted that the existing results can only handle constant input disturbances by continuous control laws  or disturbances with known {bounds} by discontinuous control laws.
The first case is a special case of our result. The second case
cannot cover our results because the bound of our disturbance is unknown. Moreover, our control law is smooth.

\end{abstract}




\section{Introduction}\label{Section I}

A formation can be defined by inter-neighbor relative positions, distances, or bearings \cite{formation2020, zhao2015translational}.
Since the  bearing-based formation is amenable to translational and scaling maneuvers and is widely used in small UAVs equipped with  onboard sensors,
the bearing-based formation control is gaining attention, see, for example,  {\cite{cao2019bearing}}, \cite{li2019multilayer}, 
 \cite{trinh2018bearing, zhao2015translational, zhao2016localizability} and the references therein.
Specifically, the localizability and distributed protocols for bearing-based network
localization was addressed in \cite{zhao2016localizability}, which lays the foundation for  the  {bearing-based} formation control.
The translational and scaling formation maneuver control was studied in \cite{zhao2015translational} via a bearing-based approach.
 Bearing-based  multilayer formation control was considered in \cite{li2019multilayer}. A bearing-ratio-of-distance rigidity theory with
application to similar formation control was established in \cite{cao2019bearing}.
The conditions and applications of the minimal and redundant bearing rigidity  were given in \cite{trinh2019minimal}.

In practice, the dynamics of the agents are often subject to some external disturbances caused by, for instance,  the  insensitivity of the sensors, uncertainty of loads.
Thus, it is interesting to further investigate the impact of the external disturbances on the performance of the bearing-based formation control law.
In fact, reference \cite{zhao2015translational} proposed a bearing-based control law that can achieve formation control in the presence of {a} unknown constant input disturbance.
Reference \cite{trinh2021robust} further considered the case where input disturbances can be any uniformly continuous and bounded time function provided that the bound of the disturbance is known.

In this paper, we further consider the problem of the bearing-based formation control with disturbance rejection for a group of agents whose dynamics, like \cite{trinh2021robust} and \cite{zhao2015translational}, are double integrators. Our disturbances are in the form of a trigonometric polynomial with arbitrary unknown amplitudes, unknown initial phases, and known or unknown frequencies, which include the constant disturbance studied  in \cite{zhao2015translational} as a special case.
For the case of
the known frequencies, we employ the canonical internal model to solve the problem, and,  for the case of
the unknown frequencies,  we combine the canonical internal model and {the} distributed adaptive control technique to deal with the problem.
It is noted that our results
cannot be covered by the results in \cite{trinh2021robust} since the bound of our disturbance is unknown.
%
%
%
%

It is noted that another topic relevant to this paper is called
 the bearing-only formation control problem, where the control of each agent can only make use of the relative bearings between itself and its neighbors and cannot make use of the relative  distances and positions  \cite{cao2021bearing},  \cite{chan2021stability}, \cite{li2018localization} - \cite{zhao2019bearing}{.} 
 The disturbance rejection problem of  the bearing-only formation control is yet to be studied.

The rest of this paper is organized as follows.
We present preliminaries and formulate the problem in Section \ref{Section Two}.
In Section \ref{Section Three}, we present the main results.
Finally, the paper is concluded in Section \ref{Section Five}.


\medskip

\noindent {\em Notation.} Let  $\otimes$ {denote} the Kronecker product. For column vectors $x_i,~i=1,\cdots,s$,  $\mbox{col} (x_1,\cdots,x_s )= [x_1^T,\cdots,x_s^T  ]^T$. $\bm{1}_{n}$ represents an $n$-dimensional vector with all elements $1$. $I_{n}$ denotes an identity matrix of size $n$. {$0_{n \times m}\in \RR^{n\times m}$ denotes a matrix with all entries $0$.}




\section{Preliminaries and Problem Formulation}\label{Section Two}

\subsection{Preliminaries}
In this section, we consider the bearing-based formation with disturbance rejection for a group of $n$ agents in $\mathbb{R}^{d}$ ($n \geq 3$ and $d \geq 2$).
Let $\mathcal{V}=\{{1},\cdots,{n}\}$ be an index set representing all agents. Partition
 $\VV=\VV_{l} \cup \VV_{f}$ with $\VV_{l}=\{1,\cdots, n_{l}\}$ denoting the index  set of leaders, and $\VV_{f}=\{n_{l}+1,\cdots, n\}$ denoting the index  set of followers with $n_{f}=n-n_{l}$ followers. Let $p_{i}\in \RR^{d}$ and $v_{i}\in \RR^{d}$ be the position and the velocity of the agent $i$, respectively,
 We assume that, for $ i \in \VV_{l}$, all agents move with a common velocity, and  for $ i \in \VV_{f}$, the {dynamics} of the  agent $i$ is described as follows:
	\begin{align}\label{sysn}
	\begin{bmatrix}
		\dot{p}_{i}\\ \dot{v}_{i}
	\end{bmatrix}=\begin{bmatrix}
		0_{d\times d} & I_{d} \\ 0_{d\times d} &0_{d\times d}
	\end{bmatrix} \begin{bmatrix}
		{p}_{i}\\ {v}_{i}
	\end{bmatrix}+\begin{bmatrix}
		0_{d\times d} \\ I_{d}
	\end{bmatrix}u_{i}+\begin{bmatrix}
		0_{d\times d} \\ I_{d}
	\end{bmatrix}d_{i}
\end{align}
where $u_{i} \in \RR^{d}$ is the input, and $d_{i}\in \RR^{d}$ is the disturbance, which is generated by some exosystem described as follows:
\begin{subequations}\label{exo}
	\begin{align}
	 \dot{\zeta}_{i}&=S_{i}{\zeta}_{i}\\
		d_{i}&=F_{i} {\zeta}_{i}\label{exo2}
	\end{align}
\end{subequations}
where ${\zeta}_{i} \in \RR^{q_{i}}$, $S_{i} \in \mathbb{R}^{q_{i}\times q_{i}}$, and  $F_{i} \in \mathbb{R}^{d\times q_{i}}$.

 Let  $p_{l}=\mbox{col}(p_{1},\cdots,p_{n_{l}})
$, $p_{f}=\mbox{col}(
	p_{n_{l}+1},\cdots,p_{n}
)$,
$v_{l}=\mbox{col}(
	v_{1},\cdots,v_{n_{l}}
)$, $v_{f}=\mbox{col}(
	v_{n_{l}+1},\cdots,v_{n}
)$,
$p=\mbox{col}(
p_{l},p_{f}
)$,
and $v=\mbox{col}(
v_{l},v_{f})$.


To describe the information exchange among different agents,  we introduce a so-called  sensing graph denoted by $\mathcal{G}\triangleq(\mathcal{V},\mathcal{E})$, where  $\mathcal{V} = \{1, 2, \cdots, n\}$ is the {index} set and $\mathcal{E}\subseteq \{({i},{j}) | {i}, {j} \in \mathcal{V}, {i} \neq {j} \}$ is an edge set with $m$ edges.   If $(i,j) \in \mathcal{E}$, we call agent $j$ a neighbor of agent $i$, and the neighbor set of agent $i$ is represented as $\mathcal{N}_{i}=\{{j} \in \mathcal{V}|(i,j)\in \mathcal{E}\}$. For an edge $(i,j)$,  index  $i$ and  index  $j$ are called the head and the tail, respectively. We assume the sensing graph $\mathcal{G}$ is undirected, i.e., for all $i,j = 1, 2, \cdots, n, {i} \neq {j}$,  $(i,j) \in \EE \Leftrightarrow (j,i) \in \EE $.
In the sensing graph, an index  ${i} \in \VV$ denotes an agent in the group and an edge $(i,j) \in \EE$ if  the control $u_i$ can make use of the
 relative  displacements  $\{p_{i}-p_{j}\}$ and the relative velocities  $\{v_{i}-v_{j}\}$.
 Let $p_{ij}\triangleq p_{i}-p_{j}$ and $v_{ij}\triangleq  v_{i}-v_{j} $.

In what follows, we use $({\cal G}, p)$ to denote a formation with the components of $\VV_l$ as the leaders that move with a common velocity and  the components of  $\VV_f$ as the followers whose {dynamics} are governed by (\ref{sysn}).

 The bearing vector $g_{ij}$ is the unit vector of $p_{ij}$ defined  as follows:
\begin{align*}
	g_{ij}\triangleq \frac{p_{ij}}{\|p_{ij} \|}.
\end{align*}
For each bearing vector $g_{ij}$, one can define
an orthogonal projector  as follows:
\begin{align*}
	\Proj_{g_{ij}}\triangleq I_{d}-g_{ij}g_{ij}^{T}.
\end{align*}
It is noted that $\Proj_{g_{ij}}$ is positive semi-definite and satisfies the idempotent property.

Let $\{g_{ij}^{*}\}_{(i,j) \in \EE}$ be the desired bearings of $\{g_{ij}\}_{(i,j) \in \EE}$.
The target formation denoted by $({\cal G}, p^*)$ is described as follows:
\begin{Definition}[Target formation]
The target formation $({\cal G}, p^*)$  satisfies the constant bearing constraints $\{g_{ij}^{*}\}_{(i,j) \in \EE}$, and the leaders move with a common constant velocity $v_{c}$.
\end{Definition}

Given a set of desired  bearings  $\{g_{ij}^{*}\}_{(i,j) \in \EE}$ and a set of leaders' positions { $p_{l}^{*}=\mbox{col}(
	p_{1}^{*},\cdots,p_{n_{l}}^{*}
)$ }  with a common constant velocity $v_{c}$, the problem of finding a set of desired followers' positions $p^*_{f}=\mbox{col}(
	p_{n_{l}+1}^{*},\cdots,p_{n}^{*}
)$ such that $({\cal G}, p^*)$ is a target formation is called the localization problem.
Such a problem may have a unique solution or multiple solutions or no solution \cite{zhao2016localizability}.


To guarantee the target formation defined above  exists and is unique, let us introduce the bearing Laplacian $\mathcal{B} \in \RR^{nd \times nd}$ associated with $\{g_{ij}^{*}\}_{(i,j) \in \EE}$
whose
$ij$th submatrix block of $\mathcal{B}$ is as follows \cite{zhao2016localizability}:
\begin{align*}
	\mathcal{B}_{ij}=\left\{ \begin{aligned}
		&0_{d\times d}  & i\neq j,&~ (i,j)\not\in \EE\\
		&-\Proj_{g_{ij}^{*}} & i\neq j,&~ (i,j)\in \EE \\
		&\sum_{j \in \mathcal{N}_{i}}\Proj_{g_{ij}^{*}} &i= j. &
	\end{aligned} \right.
\end{align*}
Partition $\mathcal{B}$ as follows:
\begin{align*}
	\mathcal{B}=\begin{bmatrix}
		B_{ll} & B_{lf}\\B_{fl} & B_{ff}
	\end{bmatrix}
\end{align*}
where $B_{ll} \in \RR^{n_{l}d\times n_{l}d}$, $B_{lf},~B_{fl}^{T}\in \RR^{n_{l}d\times n_{f}d}$, and $B_{ff} \in \RR^{n_{f}d\times n_{f}d}$.
Then, the existence and  the uniqueness condition of the target formation $(\mathcal{G}, p^{*})$ is given  as follows:
\begin{Lemma}[Theorem 1 of \cite{zhao2016localizability}]\label{unique_con}
	The target formation $(\mathcal{G}, p^{*})$ can be uniquely determined by desired bearing constraints $\{g_{ij}^{*}\}_{(i,j) \in \EE}$ and leaders' positions $p_{l}^{*}$ iff $\mathcal{B}_{ff}$ is nonsingular.
\end{Lemma}
\begin{Remark}
	According to Lemma \ref{unique_con}, if $B_{ff}$ is nonsingular, the followers' desired positions and desired velocities can be uniquely determined by $p_{f}^{*}=-B_{ff}^{-1}B_{fl}p_{l}^{*}$ and $v_{f}^{*}=-B_{ff}^{-1}B_{fl}v_{l}^{*}$, respectively. In addition, if the leaders have a common constant velocity $v_{c}\in \RR^{d}$, then $v_{f}^{*}=\bm{1}_{n_{f}} \otimes v_{c}$.
\end{Remark}

\subsection{Problem Formulation}

 Let us first describe our control law as follows: for $ i \in \VV_{f}$,
\begin{subequations}\label{cont}
\begin{align}
	u_i & = H_i (p_i, v_i, \zeta_i, \sum_{j \in \mathcal{N}_{i}} \Proj_{g_{ij}^{*}} p_{ij}, \sum_{j \in \mathcal{N}_{i}} \Proj_{g_{ij}^{*}} v_{ij}) \label{cont1} \\
\zeta_i & = G_i (p_i, v_i, \zeta_i, \sum_{j \in \mathcal{N}_{i}} \Proj_{g_{ij}^{*}} p_{ij}, \sum_{j \in \mathcal{N}_{i}} \Proj_{g_{ij}^{*}} v_{ij}) \label{cont2}
\end{align}
\end{subequations}
where $\zeta_i \in \RR^{n_c}$ for some integer $n_c$ is the state of the dynamic compensator \eqref{cont2}, $H_i$ and $G_i$ are two globally defined smooth functions. As, for each $ i \in \VV_{f}$, $u_i$  only relies on the state variables of the agent $i$ and the bearing-based orthogonal projections of the  local relative displacements $\sum_{j \in \mathcal{N}_{i}}\Proj_{g_{ij}^{*}}p_{ij}$ and the  local relative velocities
$\sum_{j \in \mathcal{N}_{i}}\Proj_{g_{ij}^{*}}v_{ij}$,  the type of control laws \eqref{cont} is called distributed  bearing-based control law.

We are now ready to {describe} our problem as follows:

\begin{Problem}\label{pro1}
Given  desired bearing constraints $\{g_{ij}^{*}\}_{(i,j) \in \EE}$ and the system (\ref{sysn}) with the disturbance generated by \eqref{exo}, assume $p_l (t) = p^*_l (t)$ for all $t \geq 0$, find a  control  law of the form \eqref{cont} such that the formation $(\mathcal{G}, p)$ converges to the target formation $(\mathcal{G}, p^{*})$ asymptotically in the sense that
$\lim_{t\to\infty} (p_{f} - p_{f}^{*}) = 0$ and {$\lim_{t\to\infty} (v_{f}-v_{f}^{*})=0$}.

\end{Problem}

Like in \cite{zhao2019bearing}, we make  the following two assumptions.

\begin{Assumption}\label{unique_assum}
$B_{ff}$ is nonsingular. 	
\end{Assumption}

\begin{Assumption}\label{collison}
	During the formation evolution, no collision occurs among agents.
\end{Assumption}
\begin{Remark}
Under Assumption \ref{unique_assum}, the target formation $(\mathcal{G}, p^{*})$  exists uniquely.
Assumption \ref{collison} excludes the collision so that the problem is well posed.
Whether or not the collision will occur depends on the initial conditions of the closed-loop system. In Section III, we will  present the sufficient conditions on the initial conditions of the closed-loop system to prevent the collision among agents.
\end{Remark}

We also need one assumption regarding the disturbances.

\begin{Assumption}\label{disturb_assum}
$\forall i \in \VV_{f}$, 	the matrices $S_{i}$ are neutrally stable.
\end{Assumption}

\begin{Remark}
 By the	neutral stability, we mean the eigenvalues of ${S}_{i}$ are  on the imaginary axis and semi-simple.
Under Assumption \ref{disturb_assum}, the disturbance $d_{i}$ can be expressed as follows:
	\begin{align}\label{disturbance}
		d_{i}=C_{i0}+\sum_{j=1}^{r_{i}}\begin{bmatrix}
			C_{ij}^{1}\sin(\omega_{ij}t+\varphi_{ij}^{1})\\ \vdots \\  C_{ij}^{d}\sin(\omega_{ij}t+\varphi_{ij}^{d})
		\end{bmatrix}
	\end{align}
	where $\omega_{i1}, \cdots, \omega_{i,r_{i}}$ are distinct positive numbers,  $C_{i0}\in \RR^{d}$, $C_{ij}^{k}\in \RR$, and $\varphi_{i1}^{1}, \cdots, \varphi_{i,r_{i}}^{d} \in \RR$ are initial phases.  Without loss of generality, we assume $C_{i0}$ is nonzero so that
the minimal polynomial of $S_{i}$ is as follows:
	\begin{align}
		\chi_{i}(\lambda)=\lambda^{2r_{i}+1}+a_{i,1}\lambda^{2r_{i}}+\cdots+a_{i,2r_{i}+1}\label{min_poly1}
	\end{align}
where $a_{i,1}, \cdots, a_{i,2r_{i}+1}\in \RR$ depends on $\mbox{col}(
		\omega_{i1},\cdots,\omega_{i,r_{i}} )$ only.
\end{Remark}

\section{Main Results}\label{Section Three}

We will consider two  scenarios, namely, the frequencies of the disturbance are known and unknown.
$\forall i \in \VV_{f}$,  let $\sigma_{i}\triangleq \mbox{col}(
		\omega_{i1},\cdots,\omega_{i,r_{i}})$.
Then, in the first case,
$\sigma_{i}$ {are} known precisely and, in the second case,  $\sigma_{i}$ {are} unknown.

\subsection{Canonical internal model}
We will use the internal model design to handle the disturbance.
According to  (\ref{disturbance}) and (\ref{min_poly1}), it can be verified that
\begin{align}
	\frac{d^{2r_{i}+1}}{dt}(d_{i})+a_{i,1}\frac{d^{2r_{i}}}{dt}(d_{i})+\cdots+a_{i,2r_{i}+1}=0\label{min_poly2}
\end{align}
where $a_{i,1}, \cdots, a_{i,2r_{i}+1}$ are functions of $\sigma_{i}$.

Let
\begin{align*}
	\Phi_{i}^{\sigma_{i}}&=\begin{bmatrix}
		0&1&\cdots&0\\
		\vdots&\vdots&\ddots&\vdots\\
			0&0&\cdots&1\\
		-a_{i,2r_{i}+1}&-a_{i,2r_{i}}&\cdots&-a_{i,1}
	\end{bmatrix}\\&\quad \in \RR^{(2r_{i}+1)\times(2r_{i}+1)}\\
	\Psi_{i}&=\begin{bmatrix}
		1&0&0&\cdots&0
	\end{bmatrix} \in \RR^{1\times (2r_{i}+1)}.
\end{align*}
Then, it can be verified that $d_{i}$ can be generated by the following exosystem:
\begin{subequations}
 \begin{align}\label{exo1}
	\dot{\vartheta}_{i}&=(\Phi_{i}^{\sigma_{i}}\otimes I_{d}){\vartheta}_{i}\\
	d_{i}&=(\Psi_{i}\otimes I_{d}) {\vartheta}_{i} \label{exo21}
\end{align}
\end{subequations}
 where $\vartheta_{i} \in \RR^{(2r_{i}+1)d}$.

Since  the pair $(\Phi_{i}^{\sigma_{i}}, \Psi_{i})$ is observable, by Proposition A.2 of \cite{huang2004nonlinear},  for any controllable pair $(M_{i}, N_{i})$ with $M_{i} \in \RR^{(2r_{i}+1)\times(2r_{i}+1)}$ Hurwitz, and $N_{i}\in \RR^{(2r_{i}+1)\times 1}$,
there exists a  unique nonsingular solution $T_{i}^{\sigma_{i}}$ to the following
Sylvester equation:
	 \begin{align}\label{Syl_eq}
	 T_{i}^{\sigma_{i}}\Phi_{i}^{\sigma_{i}}-M_{i}T_{i}^{\sigma_{i}}=N_{i}\Psi_{i}.
	 \end{align}

We now propose the following dynamic compensator:
\begin{align}\label{dy_com}
	\dot{\eta}_{i}=(M_{i}\otimes I_{d})\eta_{i}+(N_{i}\otimes I_{d})u_{i}-( M_{i}N_{i}\otimes I_{d})v_{i}
\end{align}
where $\eta_{i} \in \RR^{d(2r_{i}+1)}$, and $u_{i}$ and $v_{i}$ are introduced in \eqref{sysn}.

Performing a state transformation $\xi_{i}={\eta}_{i}+(T_{i}^{\sigma_{i}}\otimes I_{d})\vartheta_{i}-(N_{i}\otimes I_{d})v_{i}$ on (\ref{dy_com}) and making use of
(\ref{exo21}) and (\ref{Syl_eq})
{yields}
\begin{align}
	\dot{\xi}_{i}&=(M_{i}\otimes I_{d})\eta_{i}+(N_{i}\otimes I_{d})u_{i}-(M_{i}N_{i}\otimes I_{d})v_{i} \notag \\
	&\quad +(T_{i}^{\sigma_{i}}\Phi_{i}^{\sigma_{i}}\otimes I_{d})\vartheta_{i}-(N_{i}\otimes I_{d})u_{i}-(N_{i}\otimes I_{d})d_{i}\notag \\
	&=(M_{i}\otimes I_{d}){\xi}_{i}-(M_{i}T_{i}^{\sigma_{i}}\otimes I_{d})\vartheta_{i}+(M_{i}N_{i}\otimes I_{d})v_{i}\notag\\
	&\quad +(N_{i}\otimes I_{d})u_{i}-(M_{i}N_{i}\otimes I_{d})v_{i} \notag \\
	&\quad +(T_{i}^{\sigma_{i}}\Phi_{i}^{\sigma_{i}}\otimes I_{d})\vartheta_{i}-(N_{i}\otimes I_{d})u_{i}-(N_{i}\otimes I_{d})d_{i}\notag\\
	&=(M_{i}\otimes I_{d}){\xi}_{i}+((T_{i}^{\sigma_{i}}\Phi_{i}^{\sigma_{i}}-M_{i}T_{i} ^{\sigma_{i}})\otimes I_{d})\vartheta_{i}\notag\\
	&\quad -(N_{i}\Psi_{i}\otimes I_{d})\vartheta_{i}\notag\\
	&=(M_{i}\otimes I_{d}){\xi}_{i}. \label{dot_ep}
\end{align}

Substituting (\ref{exo21}) into (\ref{sysn}) gives
\begin{align}
	\begin{bmatrix}
	\dot{p}_{i}\\ \dot{v}_{i}
\end{bmatrix}&=\begin{bmatrix}
	0_{d\times d} & I_{d} \\ 0_{d\times d} &0_{d\times d}
\end{bmatrix} \begin{bmatrix}
	{p}_{i}\\ {v}_{i}
\end{bmatrix}+\begin{bmatrix}
	0_{d\times d} \\ I_{d}
\end{bmatrix}u_{i} \notag \\& \quad +
 \begin{bmatrix}
	0_{d\times d} \\ I_{d}
\end{bmatrix}(\Psi_{i}(T_{i}^{\sigma_{i}})^{-1}\otimes I_{d})(T_{i}^{\sigma_{i}}\otimes I_{d})\vartheta_{i}\notag\\
&=\begin{bmatrix}
	0_{d\times d} & I_{d} \\ 0_{d\times d} &0_{d\times d}
\end{bmatrix} \begin{bmatrix}
	{p}_{i}\\ {v}_{i}
\end{bmatrix}+\begin{bmatrix}
	0_{d\times d} \\ I_{d}
\end{bmatrix}u_{i} \notag \\& \quad +
\begin{bmatrix}
	0 _{d\times d}\\ I_{d}
\end{bmatrix}(E_{i}^{\sigma_{i}}\otimes I_{d})(T_{i}^{\sigma_{i}}\otimes I_{d})\vartheta_{i}
\label{sys_mod}
\end{align}
where $E_{i}^{\sigma_{i}}=\Psi_{i}(T_{i}^{\sigma_{i}})^{-1}$.

\begin{Remark} \label{re5}
 The dynamic compensator (\ref{dy_com}) is inspired from the results in \cite{chen2009attitude}, which is in turn based on
the internal model principle of the output regulation theory as can be found in, for example,
\cite{chen2014attitude}, \cite{huang2004nonlinear}, and the more recent survey paper \cite{huang2018}.  The key idea of the internal model approach is to convert the disturbance rejection problem of the given plant into a stabilization problem of an augmented system composed of the given plant and the internal model. A great challenge of applying the internal model principle is to find a suitable internal model that makes the  augmented system
stabilizable by the class of the prescribed control laws. 
\end{Remark}

\subsection{$\forall i \in \VV_{f}$, $\sigma_{i}$ are known}\label{known_case}
In this subsection, we consider the scenario where the frequencies of the disturbance are known.
In this case, $E_{i}^{\sigma_{i}}$ are known. We propose the control law as follows:  $\forall i \in \VV_{f}$,
\begin{subequations}\label{law_kn}
\begin{align}
	u_{i}&=(E_{i}^{\sigma_{i}}\otimes I_{d})(\eta_{i}-(N_{i}\otimes I_{d})v_{i})\notag \\
	& \quad -\kappa_{p}\sum_{j \in \mathcal{N}_{i}}\Proj_{g_{ij}^{*}}p_{ij}-\kappa_{v}\sum_{j \in \mathcal{N}_{i}}\Proj_{g_{ij}^{*}}v_{ij}\\
	\dot{\eta}_{i}&=(M_{i}\otimes I_{d})\eta_{i}+(N_{i}\otimes I_{d})u_{i}-(M_{i}N_{i}\otimes I_{d})v_{i}
\end{align}
\end{subequations}
where $\kappa_{p},~\kappa_{v}>0$.

Now, we state our first main result as follows.
\begin{Theorem}\label{Th1}
Under Assumptions \ref{unique_assum} to \ref{disturb_assum}, the control law \eqref{law_kn} is such that $\lim_{t\to\infty} (p_{f} - p_{f}^{*}) = 0$ and {$\lim_{t\to\infty} (v_{f}-v_{f}^{*})=0$}, exponentially.
\end{Theorem}
\begin{Proof}
Substituting the control law \eqref{law_kn} into \eqref{sys_mod} gives the following closed-loop system:  $\forall i \in \VV_{f}$,
\begin{subequations}\label{close_sys}
\begin{align}
\dot{p}_{i}&=v_{i} \\
\dot{v}_{i}&=(E_{i}^{\sigma_{i}}\otimes I_{d})(\eta_{i}-(N_{i}\otimes I_{d})v_{i}) \notag  -\kappa_{p}\sum_{j \in \mathcal{N}_{i}}\Proj_{g_{ij}^{*}}p_{ij}\\& \quad
-\kappa_{v}\sum_{j \in \mathcal{N}_{i}}\Proj_{g_{ij}^{*}}v_{ij}+ (E_{i}^{\sigma_{i}}\otimes I_{d})(T_{i}^{\sigma_{i}}\otimes I_{d})\vartheta_{i}  \label{close_sys2}\\
\dot{\eta}_{i}&=(M_{i}\otimes I_{d}){\eta}_{i}+(N_{i}\otimes I_{d})u_{i}-(M_{i}N_{i}\otimes I_{d})v_{i}.
\end{align}
\end{subequations}

%
Recalling the state transformation $\xi_{i}={\eta}_{i}+(T_{i}^{\sigma_{i}}\otimes I_{d})\vartheta_{i}-(N_{i}\otimes I_{d})v_{i}$ and (\ref{dot_ep}) converts (\ref{close_sys}) to the following form:
\begin{subequations}\label{close_sys1}
	\begin{align}
		\dot{p}_{i}&=v_{i} \\
		\dot{v}_{i}&=(E_{i}^{\sigma_{i}}\otimes I_{d}){\xi}_{i}\notag \\ &\quad -\kappa_{p}\sum_{j \in \mathcal{N}_{i}}\Proj_{g_{ij}^{*}}p_{ij}-\kappa_{v}\sum_{j \in \mathcal{N}_{i}}\Proj_{g_{ij}^{*}}v_{ij}\\
		\dot{\xi}_{i}	&=(M_{i}\otimes I_{d}){\xi}_{i}.
	\end{align}
\end{subequations}
Let ${\tilde{p}}_{f}=p_{f}-p_{f}^{*}$, ${\tilde{v}}_{f}=v_{f}-v_{f}^{*}$,  ${\xi}_{f}=\mbox{col}(
	{\xi}_{n_{l}+1},\cdots,{\xi}_{n}
)$,   $
\sigma_{f}= \mbox{col}(
\sigma_{n_{f}+1},\cdots,\sigma_{n}
)$, $E_{f}^{\sigma_{f}}=\mbox{block~diag }(E_{n_{l}+1}^{\sigma_{n_{l}+1}}\otimes I_{d},~\cdots,~E_{n}^{\sigma_{n}}\otimes I_{d})$, and $M_{f}=\mbox{block~diag }(M_{n_{l}+1}\otimes I_{d},~\cdots,~M_{n}\otimes I_{d})$, and use the following relations
\begin{align*}
 &\Proj_{g_{ij}^{*}}(p_{i}^{*}-p_{j}^{*})=0\\
  &\Proj_{g_{ij}^{*}}(v_{i}^{*}-v_{j}^{*})=0.
 \end{align*}

Then, we can put the closed-loop system (\ref{close_sys})  into the following compact form:
\begin{align}\label{com_sys1}
\begin{bmatrix}
\dot{\tilde{p}}_{f}\\ \dot{\tilde{v}}_{f} \\ \dot{\xi}_{f}
\end{bmatrix}=\underbrace{
\begin{bmatrix}
	\begin{array}{cc | c}
0_{dn_{f}\times dn_{f}}&I_{{dn_{f}}} &0_{dn_{f}\times q_{f}}\\
-\kappa_{p}B_{ff}&	-\kappa_{v}B_{ff}& E_{f}^{\sigma_{f}}\\ \hline
0_{q_{f}\times dn_{f}}&0_{q_{f}\times dn_{f}}&M_{f}
	\end{array}
\end{bmatrix}}_{\triangleq A_{\sigma_{f}}} \begin{bmatrix}
\tilde{p}_{f}\\ \tilde{v}_{f} \\ {\xi}_{f}
\end{bmatrix}
\end{align}
{ where $q_{f}=\sum_{i=n_{l}+1}^{n}(2r_{i}+1)d$.}

It is noted that $A_{\sigma_{f}}$ is a
block upper triangular matrix where the first diagonal block of $A_{\sigma_{f}}$ is Hurwitz for any $ \kappa_{p},\kappa_{v}>0$, and $M_{f}$ is Hurwitz by our design.
Thus, the closed-loop system  \eqref{com_sys1} is exponentially stable, which concludes the proof.
\end{Proof}

\subsection{$\forall i \in \VV_{f}$, $\sigma_{i}$ are unknown}\label{unknown_case}

In this subsection, we will further consider the scenario where the frequencies of the disturbance are unknown, which implies, $\forall i \in \VV_{f}$,
$\sigma_{i}$ and hence $E_{i}^{\sigma_{i}}$ are unknown. Thus the control law (\ref{law_kn}) does not work. To overcome this difficulty, we need to further introduce the adaptive control technique to handle the unknown parameter vectors $E_{i}^{\sigma_{i}}$. For this purpose,
let $E_{i}^{o}$ be the nominal  value of $E_{i}^{\sigma_{i}}$.
Then, $E_{i}^{\sigma_{i}}$ admit the following form:
\begin{align}\label{para}
	E_{i}^{\sigma_{i}}=E_{i}^{o }+\sum_{j=1}^{k_{i}}E_{i}^{j}\theta_{i}^{j}(\sigma_{i})
\end{align}
where $E_{i}^{o},E_{i}^{1},\cdots,E_{i}^{k_{i}}\in \mathbb{R}^{1 \times (2r_{i}+1)}$ are known matrices, and $\theta_{i}^{1}(\cdot),~\cdots,~ \theta_{i}^{k_{i}}(\cdot) \in \RR$ are known functions of $\sigma_i$.
Let
$\hat{\theta}_{i}^{1} ,~\cdots,~\hat{\theta}_{i}^{k_{i}}\in \RR$ be the estimates of ${\theta}_{i}^{1}(\sigma_{i}),~\cdots,~{\theta}_{i}^{k_{i}}(\sigma_{i})$, respectively, and let
$\theta_{i}(\sigma_{i})=\mbox{col}(
		{\theta}_{i}^{1}(\sigma_{i}),~\cdots,~{\theta}_{i}^{k_{i}}(\sigma_{i})
)$, $\hat{\theta}_{i}=\mbox{col}(
	\hat{\theta}_{i}^{1},~\cdots,~\hat{\theta}_{i}^{k_{i}}
)$, $\hat{E}_{i}=E_{i}^{o}+\sum_{j=1}^{k_{i}}E_{i}^{j}\hat{\theta}_{i}^{j}$.
Then, we propose the following adaptive control law:  $\forall i \in \VV_{f}$,
\begin{subequations}\label{law_unkn}
	\begin{align}
		u_{i}&=(\hat{E}_{i}\otimes I_{d})(\eta_{i}-(N_{i}\otimes I_{d})v_{i})\notag \\& \quad -\kappa_{p}\sum_{j \in \mathcal{N}_{i}}\Proj_{g_{ij}^{*}}p_{ij}-\kappa_{v} \sum_{j \in \mathcal{N}_{i}}\Proj_{g_{ij}^{*}}v_{ij} \label{law_unkn1}\\
		\dot{\eta}_{i}&=(M_{i}\otimes I_{d}){\eta}_{i}+(N_{i}\otimes I_{d})u_{i}-(M_{i}N_{i}\otimes I_{d})v_{i} \\
		\dot{\hat{\theta}}_{i}&=-\Lambda_{i}\rho_{i}^{T}(\eta_{i}, v_{i})(\sum_{j \in \mathcal{N}_{i}}\Proj_{g_{ij}^{*}}p_{ij}+\sum_{j \in \mathcal{N}_{i}}\Proj_{g_{ij}^{*}}v_{ij})
	\end{align}
\end{subequations}
where  $\kappa_{p}>0$,  $\kappa_{v}>\frac{1}{\lambda_{min}(B_{ff})}$, $\Lambda_{i}$ are some positive definite matrices, and
\begin{align} \label{rhoi}
	&\rho_{i}(\eta_{i}, v_{i})= \notag\\
	&[
	(E_{i}^{1} \otimes I_{d})(\eta_{i}-(N_{i}\otimes I_{d})v_{i}) \cdots (E_{i}^{k_i} \otimes I_{d})(\eta_{i}-(N_{i}\otimes I_{d})v_{i})]	{.}
	\end{align}

Substituting (\ref{law_unkn1}) into  (\ref{sysn}) gives
\begin{align}
		\dot{v}_{i}&=(\hat{E}_{i}\otimes I_{d})(\eta_{i}-(N_{i}\otimes I_{d})v_{i})\notag \\& \quad -\kappa_{p}\sum_{j \in \mathcal{N}_{i}}\Proj_{g_{ij}^{*}}p_{ij}-\kappa_{v} \sum_{j \in \mathcal{N}_{i}}\Proj_{g_{ij}^{*}}v_{ij}+d_{i} .\label{close_sys_unkn2}
	\end{align}

Let $\tilde{E}_{i}=E_{i}^{\sigma_{i}}-\hat{E}_{i}$. Then  (\ref{close_sys_unkn2}) can be further rewritten  as follows:
\begin{align*}
\dot{v}_{i}&=({E}_{i}^{\sigma_{i}}\otimes I_{d})(\eta_{i}-(N_{i}\otimes I_{d})v_{i})\\
&\quad -(\tilde{E}_{i}\otimes I_{d})(\eta_{i}-(N_{i}\otimes I_{d})v_{i}) \\
& \quad -\kappa_{p}\sum_{j \in \mathcal{N}_{i}}\Proj_{g_{ij}^{*}}p_{ij}-\kappa_{v} \sum_{j \in \mathcal{N}_{i}}\Proj_{g_{ij}^{*}}v_{ij}\\
&\quad +(E_{i}^{\sigma_{i}}\otimes I_{d})(T_{i}^{\sigma_{i}}\otimes I_{d})\vartheta_{i}\\
&=({E}_{i}^{\sigma_{i}}\otimes I_{d})(\eta_{i}+(T_{i}^{\sigma_{i}}\otimes I_{d})\vartheta_{i}-(N_{i}\otimes I_{d})v_{i}))\\
&\quad -(\tilde{E}_{i}\otimes I_{d})(\eta_{i}-(N_{i}\otimes I_{d})v_{i}) \\
& \quad -\kappa_{p}\sum_{j \in \mathcal{N}_{i}}\Proj_{g_{ij}^{*}}p_{ij}-\kappa_{v} \sum_{j \in \mathcal{N}_{i}}\Proj_{g_{ij}^{*}}v_{ij}\\
&=(E_{i}^{\sigma_{i}}\otimes I_{d}){\xi}_{i}-(\tilde{E}_{i}\otimes I_{d})(\eta_{i}-(N_{i}\otimes I_{d})v_{i})\notag \\ &\quad -\kappa_{p}\sum_{j \in \mathcal{N}_{i}}\Proj_{g_{ij}^{*}}p_{ij}-\kappa_{v}\sum_{j \in \mathcal{N}_{i}}\Proj_{g_{ij}^{*}}v_{ij}.
\end{align*}

Let $\tilde{\theta}_{i}^{j}={\theta}_{i}^{j}(\sigma_{i})-\hat{\theta}_{i}^{j}$ and $\tilde{\theta}_{i}=\mbox{col}(
	\tilde{\theta}_{i}^{1},\cdots,\tilde{\theta}_{i}^{{k_{i}}}
)$.  Then,  we have
\begin{align*}
&\quad ~(\tilde{E}_{i}\otimes I_{d})(\eta_{i}-(N_{i}\otimes I_{d})v_{i})\\
&=(\sum_{j=1}^{k_{i}}E_{i}^{j}\tilde{\theta}_{i}^{j} \otimes I_{d})(\eta_{i}-(N_{i}\otimes I_{d})v_{i})\\
&=\sum_{j=1}^{k_{i}}(E_{i}^{j} \otimes I_{d})(\eta_{i}-(N_{i}\otimes I_{d})v_{i}) \tilde{\theta}_{i}^{j}\\
&=\rho_{i}(\eta_{i}, v_{i})\tilde{\theta}_{i}.
\end{align*}

Thus, we have
\begin{align}
\dot{v}_{i}&=(E_{i}^{\sigma_{i}}\otimes I_{d}){\xi}_{i}- \rho_{i}(\eta_{i}, v_{i})\tilde{\theta}_{i} \notag \\ &\quad -\kappa_{p}\sum_{j \in \mathcal{N}_{i}}\Proj_{g_{ij}^{*}}p_{ij}-\kappa_{v}\sum_{j \in \mathcal{N}_{i}}\Proj_{g_{ij}^{*}}v_{ij}. \label{close_sys1_unkn2x}
		\end{align}
Thus, under the control law (\ref{law_unkn}), the closed-loop system takes the following form:
\begin{subequations}\label{close_sys1_unkn}
		\begin{align}
			\dot{p}_{i}&=v_{i} \\
			\dot{v}_{i}&=(E_{i}^{\sigma_{i}}\otimes I_{d}){\xi}_{i}-  \rho_{i}(\eta_{i}, v_{i})\tilde{\theta}_{i}\notag \\ &\quad -\kappa_{p}\sum_{j \in \mathcal{N}_{i}}\Proj_{g_{ij}^{*}}p_{ij}-\kappa_{v}\sum_{j \in \mathcal{N}_{i}}\Proj_{g_{ij}^{*}}v_{ij} \label{close_sys1_unkn2}\\
			\dot{\xi}_{i}	&=(M_{i}\otimes I_{d}){\xi}_{i}\\
			\dot{\hat{\theta}}_{i}&=-\Lambda_{i}\rho_{i}^{T}(\eta_{i}, v_{i})(\sum_{j \in \mathcal{N}_{i}}\Proj_{g_{ij}^{*}}p_{ij}+\sum_{j \in \mathcal{N}_{i}}\Proj_{g_{ij}^{*}}v_{ij}).
		\end{align}
\end{subequations}


Let
\begin{align*}
	{\theta}_{f}=&\mbox{col}(
		{\theta}_{n_{l}+1}(\sigma_{i}),\cdots,{\theta}_{n}(\sigma_{i})
	) \\
\tilde{\theta}_{f}=&\mbox{col}(
		\tilde{\theta}_{n_{l}+1},\cdots,\tilde{\theta}_{n}
	)\\
    \hat{\theta}_{f}=&\mbox{col}(
	\hat{\theta}_{n_{l}+1},\cdots,\hat{\theta}_{n}
    )
    \\ \eta_{f}=&\mbox{col}(
	\eta_{n_{l}+1},\cdots,\eta_{n}
)\\
\Lambda_{f}=&\mbox{block~diag}(\Lambda_{n_{l}+1},\cdots,\Lambda_{n}),
\end{align*}
and
\begin{align*}
	&\rho_{f}(\eta_{f}, v_{f})=\\
	&\mbox{block~diag} (\rho_{n_{l}+1}(\eta_{n_{l}+1}, v_{n_{l}+1}),
	~\cdots,~\rho_{n}(\eta_{n}, v_{n})).
\end{align*}

Furthermore, noting $\dot{\hat{\theta}}_{f} = - \dot{\tilde{\theta}}_{f}$ and
\begin{align*}
	&\Proj_{g_{ij}^{*}}(p_{i}^{*}-p_{j}^{*})=0\\
	&\Proj_{g_{ij}^{*}}(v_{i}^{*}-v_{j}^{*})=0
\end{align*}
gives the compact form of \eqref{close_sys1_unkn} as follows:
\begin{subequations}\label{whole_sys_un}
	\begin{align}
		\dot{\tilde{p}}_{f}&={\tilde{v}}_{f} \\
		\dot{\tilde{v}}_{f}&=-\kappa_{p} B_{ff} {\tilde{p}}_{f}-\kappa_{v} B_{ff} {\tilde{v}}_{f}
	 +E_{f}^{\sigma_{f}}{\xi}_{f}\notag \\
	 &\quad -\rho_{f}(\eta_{f}, v_{f}) \tilde{\theta}_{f}\\
	\dot{{\xi}}_{f}&=M_{f}{{\xi}}_{f}\\
		\dot{\tilde{\theta}}_{f}&=	\Lambda_{f}\rho_{f}^{T}(\eta_{f}, v_{f})B_{ff}({\tilde{p}}_{f}+{\tilde{v}}_{f}).
	\end{align}
\end{subequations}

For convenience, let
\begin{align*}
 	\tilde{x}_{f}&=\mbox{col}(
 		\tilde{p}_{f},\tilde{v}_{f}
 	)\\
    A_{c}&=\begin{bmatrix}
    	0_{dn_{f}\times dn_{f}}&I_{dn_{f}}\\-\kappa_{p} B_{ff}& -\kappa_{v} B_{ff}
    \end{bmatrix} \\
    B_{c}&=\mbox{col}(0_{dn_{f}\times dn_{f}},I_{dn_{f}} ).
\end{align*}
Then, \eqref{whole_sys_un} can be put in the following form:
\begin{subequations}\label{whole_sys_unx}
	\begin{align}
		\tilde{x}_{f} &= A_c \tilde{x}_{f} + B_{c} (E_{f}^{\sigma_{f}}{\xi}_{f}-\rho_{f}(\eta_{f}, v_{f}) \tilde{\theta}_{f})\\
		\dot{{\xi}}_{f}&=M_{f}{{\xi}}_{f}\\
		\dot{\tilde{\theta}}_{f}&=\Lambda_{f}\rho_{f}^{T}(\eta_{f}, v_{f})B_{ff}({\tilde{p}}_{f}+{\tilde{v}}_{f}). \label{whole_sys_unxc}
	\end{align}
\end{subequations}

We now state  the following result:
\begin{Theorem}\label{Th2}
Under Assumptions \ref{unique_assum} to \ref{disturb_assum}, for any initial {conditions}, the solution of the {closed-loop} system (\ref{whole_sys_unx}) is bounded, and is such that $\lim_{t\to\infty} (p_{f} - p_{f}^{*}) = 0$ and $\lim_{t\to\infty} (v_{f} - v_{f}^{*}) = 0$.
\end{Theorem}

\begin{Proof}

Let
\begin{align}\label{qc}
Q_{c}=\mbox{block~diag}(2\kappa_{p} B_{ff}^{2}, 2(\kappa_{v} B_{ff}^{2}-B_{ff})),
\end{align}
and
\begin{align}\label{pc1}
P_{c}&=\begin{bmatrix}
	(\kappa_{p}+\kappa_{v})B_{ff}^2 & B_{ff}\\ B_{ff} &  B_{ff}
\end{bmatrix}.
\end{align}
Then it can be verified that
$Q_{c}$ is positive definite since
 \begin{align*}
 		&2\kappa_{p} B_{ff}^{2}>0\\
 		&2(\kappa_{v} B_{ff}^{2}-B_{ff})=2B_{ff}^{\frac{1}{2}}(\kappa_{v}B_{ff}-1)B_{ff}^{\frac{1}{2}}>0,
\end{align*}
and $P_c$ satisfies
\begin{align}
	A_{c}^{T}P_{c}+P_{c}A_{c}=-Q_{c}.\label{Ly_eq1}
\end{align}

Since,  for any  $\kappa_{p}>0$ and $\kappa_{v}>\frac{1}{\lambda_{min}(B_{ff})}$, $A_c$ is Hurwitz, $P_c$ must also be  positive definite.

Since $M_f$ is Hurwitz, there {exists} a unique positive definite and symmetric matrix $G_{c}\in \Sm^{\sum_{i=n_{l}+1}^{n}(2r_{i}+1) d}$ satisfying
\begin{align*}
	G_{c}M_{f}+M_{f}^{T}G_{c}=-I_{\sum_{i=n_{l}+1}^{n}(2r_{i}+1)d}.
\end{align*}
Consider the Lyapunov function candidate for (\ref{whole_sys_unx}) as follows:
\begin{align}\label{Ly1}
	V=\tilde{x}_{f}^{T}P_{c}\tilde{x}_{f}+\gamma \xi_{f}^{T}G_{c} \xi_{f}+ \tilde{\theta}_{f}^{T}\Lambda_{f}^{-1}\tilde{\theta}_{f}.
\end{align}
Taking the time derivative of (\ref{Ly1}) along the trajectory of (\ref{whole_sys_unx}) gives
\begin{align}
	\dot{V}&=\tilde{x}_{f}^{T}P_{c}(A_{c}\tilde{x}_{f}+B_{c}E_{f}^{\sigma_{f}}{\xi}_{f}-B_{c}\rho_{f}(\eta_{f}, v_{f})\tilde{\theta}_{f})\notag \\
	&\quad+(A_{c}\tilde{x}_{f}+B_{c}E_{f}^{\sigma_{f}}{\xi}_{f}-B_{c}\rho_{f}(\eta_{f}, v_{f})\tilde{\theta}_{f})^{T}P_{c}\tilde{x}_{f}\notag\\
	&\quad+\gamma\xi_{f}^{T}(G_{c}M_{f}+M_{f}^{T}G_{c})\xi_{f}\notag\\
	&\quad +2\tilde{\theta}_{f}^{T}\rho_{f}^{T}(\eta_{f}, v_{f})B_{ff}({\tilde{p}}_{f}+{\tilde{v}}_{f}) \notag\\
	&=-\tilde{x}_{f}^{T}Q_{c}\tilde{x}_{f}-\gamma \xi_{f}^{T}\xi_{f}+2\tilde{x}_{f}^{T}P_{c}B_{c}E_{f}^{\sigma_{f}}\xi_{f}\notag\\
	&\quad -2\tilde{x}_{f}^{T}P_{c}B_{c}\rho_{f}(\eta_{f}, v_{f})\tilde{\theta}_{f}
	 \notag\\&\quad +2\tilde{\theta}_{f}^{T}\rho_{f}^{T}(\eta_{f}, v_{f})B_{ff}({\tilde{p}}_{f}+{\tilde{v}}_{f}) \notag \\
	&=-\tilde{x}_{f}^{T}Q_{c}\tilde{x}_{f}-\gamma \xi_{f}^{T}\xi_{f}+2{x}_{f}^{T}P_{c}B_{c}E_{f}^{\sigma_{f}}\xi_{f} \notag\\
	& = -  \begin{bmatrix}
	\tilde{x}_{f}^T & \xi_{f}^T
\end{bmatrix}
\Omega
 \begin{bmatrix}
	\tilde{x}_{f} \\
 \xi_{f}
\end{bmatrix} \label{der_Ly1}
\end{align}
  where $ \Omega = \begin{bmatrix}
	Q_{c} &  P_{c}B_{c}E_{f}^{\sigma_{f}} \\
(P_{c}B_{c}E_{f}^{\sigma_{f}})^T & {\gamma} I
\end{bmatrix}.$
By Schur complement, $ \Omega$ is positive definite if and only if $Q_c - \frac{1}{\gamma} P_{c}B_{c}E_{f}^{\sigma_{f}} (P_{c}B_{c}E_{f}^{\sigma_{f}})^T$ is positive definite.
For any $\sigma_{f}$, let $\gamma_{\sigma_{f}}=\frac{\lambda_{max} (P_{c}B_{c}E_{f}^{\sigma_{f}} (P_{c}B_{c}E_{f}^{\sigma_{f}})^T)}{\lambda_{min} (Q_c)}$ and $\gamma>\gamma_{\sigma_{f}}$. Then, $Q_c - \frac{1}{\gamma} P_{c}B_{c}E_{f}^{\sigma_{f}} (P_{c}B_{c}E_{f}^{\sigma_{f}})^T$ is positive definite.
Thus, By LaSalle-Yoshizawa Theorem \cite{kkk95}, the solution of the closed-loop system is bounded and is such that
$\lim_{t\to\infty} (p_{f}- p_{f}^{*}) = 0$ and $\lim_{t\to\infty} (v_{f}-v_{f}^{*}) = 0$.
\end{Proof}



\section{Conclusion}\label{Section Five}
This paper has studied the problem  of the bearing-based formation control with disturbance rejection for a group of agents governed by double integrators.
The disturbances are in the form of a trigonometric polynomial with arbitrary unknown amplitudes, unknown initial phases, and known or unknown frequencies. For the case of
the known frequencies, we have employed the canonical internal model to solve the problem, and,  for the case of
the unknown frequencies,  we { have combined} the canonical internal model and {the} distributed adaptive control technique to deal with the problem.

\end{document}